\documentclass[aps,prd,twocolumn,showpacs,groupedaddress]{revtex4-1}

\usepackage{graphicx}
\usepackage{dcolumn}
\usepackage{amsmath}
\usepackage{amssymb}
\usepackage{xcolor}
\usepackage{multirow}
\usepackage{listings}
\usepackage{xcolor}
\usepackage{lineno}
\usepackage{extarrows}
\usepackage[colorlinks,linkcolor=blue,anchorcolor=blue,citecolor=blue]{hyperref}
\allowdisplaybreaks
\newcommand\dx{{\rm d}}
\newcommand\p{\partial}
\newcommand\etal{{\it et~al.}}

\begin{document}

\title{Scalar-tensor nonlocal gravity}
\author{Shuxun Tian}
\email[]{tshuxun@whu.edu.cn}
\affiliation{School of Physics and Technology, Wuhan University, 430072, Wuhan, China}
\date{\today}
\begin{abstract}
  In order to explain the late-time cosmological acceleration, we propose a new pure geometric gravity based on Deser-Woodard theory [\href{http://dx.doi.org/10.1103/PhysRevLett.99.111301}{Phys. Rev. Lett. 99, 111301 (2007)}], in which the expansion of a matter-dominated universe can be accelerating. This new theory behaves as well as general relativity in solar system and gravitational waves, which is an important improvement over Deser-Woodard nonlocal gravity. Especially, for the simplest case, theoretical considerations determine all the parameters that appear in the Lagrangian.
\end{abstract}
\maketitle

\section{Introduction}
Which theory is the most beautiful one to explain the late-time cosmological acceleration? Dark energy, e.g., quintessence \cite{Caldwell1998} or phantom \cite{Caldwell2002}, is a simple model. However, this mysterious matter is undetectable in the local gravitational systems, and thus, its existence lacks direct verification. Modified gravity is also a good candidate (see \cite{Capozziello2011,Nojiri2011,Clifton2012,Nojiri2017} for reviews). One can couple new fields to the Einstein-Hilbert Lagrangian or add higher-order geometric terms to the action. But current mainstream modified gravities are not good enough to replace Einstein's theory. The problems come from observations and aesthetics. Many simple models, e.g., $f(R)=R+\alpha R^2$ theory, can be directly ruled out by observations \cite{Clifton2012}. In order to fit all known observations, one has to construct complex structures for the Lagrangian, e.g., the Hu-Sawicki model \cite{Hu2007}. This always makes the theory not so charming, even though it fits the data well. Another aesthetics problem is that, in order to explain the late-time acceleration, classical theories generally need a parameter related to the Hubble constant $H_0$. This may be a potential factor that makes theories suffer from the coincidence problem. We believe if a pure geometric gravity does not introduce any $H_0$-related parameters into the Lagrangian and can behave well in various gravitational applications, then it is comparable to previous theories.

Pure fourth-order gravity \cite{Tian2018} seems to be a good choice. However, combined with Eq. (8) and Eq. (27) in \cite{Tian2018} and $\rho_0=\mathcal{O}(H_0^2/G)$ from dimensional analysis, we obtain $\sigma_i=\mathcal{O}(c/H_0)$, which is inconsistent with the explanation of \cite{Tian2018} that $\sigma_i$ is the size of fundamental particles. Deser-Woodard nonlocal gravity \cite{Deser2007} is a pure geometric theory and only introduces dimensionless parameters in the Lagrangian. This theory could explain the late-time acceleration, but gives $\Psi\neq\Phi$ in the weak field limit [see metric (\ref{eq:04}) for the meaning of $\Phi$ and $\Psi$] for the general formalism \cite{Koivisto2008}. The Cassini mission gives $\Psi/\Phi=1+(2.1\pm2.3)\times10^{-5}$ in the Solar system \cite{Bertotti2003}, and recently, $\Psi=\Phi$ has been confirmed at the Galaxy scale \cite{Collett2018}. We believe a good gravity theory should give $\Psi$ that exactly equals, not close, to $\Phi$ \footnote{Similar to $f(R)=R+f_1R^2+f_2R^3$ theory, setting $f_1=0$ in \cite{Koivisto2008} is not a good way to achieve this goal. Otherwise, one has to analyze the weak field limit on a more complex background, e.g., a de Sitter metric.}. One thing should be mentioned here, the Deser-Woodard nonlocal gravity could suppress the growth rate of a large scale structure formation \cite{Nersisyan2017b,Park2018}, which is in better agreement with observations, and this good property may be related to the fact that $\Psi\neq\Phi$ in the Deser-Woodard theory. If we require $\Psi=\Phi$, we may lose this merit. However, $\Psi\neq\Phi$ in modified gravities may not be the necessary condition to suppress the growth rate. Generally, the evolution equations of cosmological scalar perturbation in modified gravities and general relativity should be different. It is reasonable that different evolution equations give different growth rates. In this paper, we focus on constructing a gravity theory that not only explains the late-time acceleration, but also gives $\Psi=\Phi$ in the weak field limit. Detailed analysis of the large scale structure formation for the new theory will be presented in future works.

To our knowledge, the $\Psi=\Phi$ issue has not been discussed in the follow-up works on the Deser-Woodard nonlocal gravity (recent developments can be found in \cite{Deser2013,Maggiore2014,Nersisyan2016,Nersisyan2017a,Nersisyan2017b,Park2018,Vardanyan2018,Narain2018,Kumar2018} and references therein). In this paper, similar to the pure fourth-order gravity, we will construct a new nonlocal theory that makes $\Psi=\Phi$ through the combination of scalar and tensor modifications. In addition, we will present comprehensive calculations on the weak field limit, gravitational waves, and cosmology for the new theory. We take same the conventions as \cite{Tian2018}.

\section{Field equations}
The action takes the form
\begin{equation}
  S=\int\dx^4x\sqrt{-g}\mathcal{L}_G+S_{\rm m}.
\end{equation}
where $\delta S_{\rm m}=-\int\dx^4x\sqrt{-g}T_{\mu\nu}\delta g^{\mu\nu}/2$ and $T_{\mu\nu}=(\rho+p/c^2)u_\mu u_\nu+pg_{\mu\nu}$ for the perfect fluid. We start from the following Lagrangian:
\begin{gather}
  \mathcal{L}_G=\frac{1}{2\zeta}\left(\alpha_1R+\alpha_2R\Box^{-1}R+\alpha_3R_{\mu\nu}\Box^{-1}R^{\mu\nu}\right),
\end{gather}
where $\zeta$ is the spacetime-matter coupling constant, $\alpha_i$ is dimensionless parameter, and $\Box$ is d'Alembert operator. Without loss of generality, we can fix one parameter. However, in order to facilitate the discussions on completely abandoning Einstein-Hilbert action, we restore all $\alpha_i$. Note that the above Lagrangian can be regarded as a special case of \cite{Nersisyan2017a}. However, our motivations and work details are completely different. $\Psi=\Phi$ is our core, and this is not mentioned in \cite{Nersisyan2017a}. In addition, \cite{Nersisyan2017a} shows that a general tensor nonlocal modification will lead to an instability problem in the evolution of the Universe. But the tensor part $R_{\mu\nu}\Box^{-1}R^{\mu\nu}$ that we are considering does not bring the instability problem to the theory \cite{Nersisyan2017a}. In the Lagrangian, $R_{\mu\nu\rho\lambda}\Box^{-1}R^{\mu\nu\rho\lambda}$ is unnecessary because of Eq. (6) in \cite{Nersisyan2017a}. Variation of the action with respect to the metric gives the field equations
\begin{widetext}
\begin{align}\label{eq:03}
  &\alpha_1G_{\mu\nu}+\alpha_2\left[2g_{\mu\nu}R-2\nabla_\mu\nabla_\nu U+2UG_{\mu\nu}+(\nabla_\mu U)(\nabla_\nu U)-\frac{g_{\mu\nu}}{2}(\nabla^\lambda U)(\nabla_\lambda U)\right]\nonumber\\
  &+\alpha_3\left[R_{\mu\nu}+g_{\mu\nu}\nabla_\alpha\nabla_\beta U^{\alpha\beta}
  -\nabla_\mu\nabla_\alpha U^{\alpha}_{\ \nu}
  -\nabla_\nu\nabla_\alpha U^{\alpha}_{\ \mu}
  +2U^{\alpha\beta}R_{\mu\alpha\nu\beta}
  -\frac{g_{\mu\nu}}{4}\Box(U_{\alpha\beta}U^{\alpha\beta})
  -\frac{g_{\mu\nu}}{2}R_{\alpha\beta}U^{\alpha\beta}\right.\nonumber\\
  &\left.+(\nabla_\nu U_{\alpha\beta})(\nabla_\mu U^{\alpha\beta})
  +\nabla_\alpha(U^{\alpha\beta}\nabla_\nu U_{\mu\beta}+U^{\alpha\beta}\nabla_\mu U_{\nu\beta}
  -U_{\mu\beta}\nabla_\nu U^{\alpha\beta}-U_{\nu\beta}\nabla_\mu U^{\alpha\beta})\right]
  =\zeta T_{\mu\nu},
\end{align}
\end{widetext}
where $U_{\mu\nu}=\Box^{-1}R_{\mu\nu}$, or in a localized form $\Box U_{\mu\nu}=R_{\mu\nu}$, and $U=U^\mu_{\ \mu}$. One thing should be mentioned here, $U_{\mu\nu}$ is not a new tensor field, but a nonlocal form of the spacetime geometry. Energy and momentum conservation can be directly derived from Eq. (\ref{eq:03}).

\section{Meeting observations}
As a primary requirement, the new scalar-tensor nonlocal gravity should be able to explain the main phenomena in the Solar system (Newton's gravity and $\Psi=\Phi$), gravitational waves (speed), and cosmology (the late-time acceleration).

\subsection{Weak field limit}
A metric gravity theory should recover Newton's law of gravitation in the weak field approximation \footnote{In this paper, we do not consider MOND [M. Milgrom, \href{http://dx.doi.org/10.1086/161130}{Astrophys. J. {\bf 270}, 365 (1983)}] things.}. Furthermore, considering the observational constraints \cite{Bertotti2003,Collett2018}, we demand that the theory gives $\Psi=\Phi$ for the following perturbed line element:
\begin{equation}\label{eq:04}
  \dx s^2=-c^2(1+2\Phi/c^2)\dx t^2+(1-2\Psi/c^2)\dx\mathbf{r}^2.
\end{equation}
The background is the Minkowski metric, in which $R_{\mu\nu}=0$. Therefore, we can choose $U_{\mu\nu}=0$ for the background. In the weak field limit, i.e., in metric (\ref{eq:04}), $\Phi$ and $\Psi$ are first-order infinitesimals. Omitting the second-order infinitesimals, we obtain $R_{00}=\nabla^2\Phi$ and so on for metric (\ref{eq:04}), where $\nabla^2$ is the Laplace operator. Taking into account $\Box U_{\mu\nu}=R_{\mu\nu}$, we know $R_{\mu\nu}$ and $U_{\mu\nu}$ should be regarded as first-order infinitesimals in the weak field limit. This makes terms like $UG_{\mu\nu}$ to be second-order infinitesimals and can be ignored in Eq. (\ref{eq:03}). So the field equations can be written as $\alpha_1G_{\mu\nu}+\alpha_2\left[2g_{\mu\nu}R-2\nabla_\mu\nabla_\nu U\right]
+\alpha_3\left[R_{\mu\nu}+g_{\mu\nu}U_0-\nabla_\mu\nabla_\alpha U^{\alpha}_{\ \nu}-\nabla_\nu\nabla_\alpha U^{\alpha}_{\ \mu}\right]=\zeta T_{\mu\nu}$, where $U_0=\nabla_\alpha\nabla_\beta U^{\alpha\beta}$.

Contraction of the field equations gives
\begin{equation}\label{eq:05}
  -\alpha_1R+6\alpha_2R+\alpha_3(R+2U_0)=\zeta T,
\end{equation}
where we have used $\Box U=R$. Because Christoffel symbols vanish at the Minkowski background, we have $\nabla_\mu({\rm tensor})=\p_\mu({\rm tensor})$ except for the metric tensor $g_{\mu\nu}$. The source is assumed to be static \cite{Tian2018}, which implies $\nabla_0({\rm function})=\p_0({\rm function})=0$ for the left side of field equations. Thus, the $00$ component gives
\begin{equation}\label{eq:06}
  \alpha_1G_{00}+2\alpha_2\bar{g}_{00}R+\alpha_3(R_{00}+\bar{g}_{00}U_0)=\zeta T_{00},
\end{equation}
where a bar means background. In addition, $\nabla_\mu({\rm tensor})=\p_\mu({\rm tensor})$ means the covariant derivative indices are commutable, i.e., $\nabla_\mu\nabla_\nu({\rm tensor})=\nabla_\nu\nabla_\mu({\rm tensor})$. Applying this to $\Box U^{\mu\nu}=R^{\mu\nu}$, we obtain $\nabla_\mu\nabla_\nu R^{\mu\nu}=\nabla_\mu\nabla_\nu\Box U^{\mu\nu}=\Box(\nabla_\mu\nabla_\nu U^{\mu\nu})=\Box U_0$. The Bianchi identity gives $\nabla_\nu R^{\mu\nu}=\nabla^\mu R/2$, and then $\nabla_\mu\nabla_\nu R^{\mu\nu}=\Box R/2$. So, $\Box(U_0-R/2)=0$. In the static case, the general spherically symmetric solution for $\Box({\rm function})=0$ is $c_1+c_2/r$, where $c_i$ is constant. We expect the mass element has negligible effect on the areas far away from it. This boundary condition at infinity, which has been widely used to eliminate the increasing mode in the classical massive gravity's Newtonian approximation \cite{Clifton2012}, gives $c_1=0$ [note that a nonzero $c_1$ means a term almost similar to the cosmological constant appears in Eq. (\ref{eq:06})]. As we analyze the weak field limit, all functions should be analytical in the whole space. This condition requires $c_2=0$ \cite{Tian2018}. Then we obtain
\begin{equation}\label{eq:07}
  U_0=R/2.
\end{equation}
For the source, the only nonzero energy-momentum tensor is $T_{00}=\rho c^4$, and the trace $T=-\rho c^2$ \cite{Tian2018}.

Substituting metric (\ref{eq:04}) into Eqs. (\ref{eq:05})--(\ref{eq:07}), and eliminating $\rho$ and $U_0$, we obtain
\begin{equation}
  (\alpha_1-8\alpha_2-3\alpha_3)\nabla^2\Psi=(\alpha_1-4\alpha_2-\alpha_3)\nabla^2\Phi.
\end{equation}
If $2\alpha_2+\alpha_3=0$, then $\nabla^2\Psi=\nabla^2\Phi$, i.e., $\Psi=\Phi+c_1+c_2/r$ for the spherically symmetric case. The boundary conditions at infinity (perturbations vanish at infinity) and $r=0$ (perturbations are nondivergent in the whole space) require $c_i=0$. So, $2\alpha_2+\alpha_3=0$ gives $\Psi=\Phi$ in the weak field limit. This result also means $\Psi\neq\Phi$ in the general Deser-Woodard nonlocal gravity. Substituting $\alpha_2=-\alpha_3/2$ and $\Psi=\Phi$ into Eq. (\ref{eq:06}), we obtain
\begin{equation}
  2(\alpha_1+\alpha_3)\nabla^2\Phi=\zeta\rho c^4.
\end{equation}
Comparing with Poisson's equation $\nabla^2\Phi=4\pi G\rho$, we obtain $\zeta=8(\alpha_1+\alpha_3)\pi G/c^4$. For the case $\{\alpha_1=0,\alpha_2=-1/2,\alpha_3=1\}$, $\zeta=8\pi G/c^4$. In this case that completely abandons the Einstein-Hilbert action, there are no unknown parameters appearing in the Lagrangian. As we soon see, this simple case also behaves well in gravitational waves and cosmology.

\subsection{Gravitational waves}
Here we focus on the speed of gravitational waves. Observations give tight bounds on the graviton mass and the absolute speed \cite{GW150914,GW170817}. We expect that the new theory gives the speed of gravitational waves equal to $c$. Similar to the $f(R)=R+\alpha R^2$ theory \cite{Liang2017}, the contraction of the vacuum field equations is enough for this analysis. For simplicity, we choose the background to be a Minkowski metric. But the boundary condition at infinity used in the weak field limit analysis is unavailable here, and thus, Eq. (\ref{eq:07}) is unavailable. What we can obtain is just $\Box(U_0-R/2)=0$. Applying $\Box$ to Eq. (\ref{eq:05}), and setting $T=0$ and $\alpha_2=-\alpha_3/2$, we obtain
\begin{equation}
  (\alpha_1+\alpha_3)\Box R=0.
\end{equation}
This means no massive polarization exists and the speed of gravitational waves equal to $c$ in the new theory.

\subsection{Cosmology}
Here we focus on the flat homogeneous isotropic universe, which can be described by the flat Friedmann-Lema\^{i}tre-Robertson-Walker metric,
\begin{equation}\label{eq:11}
  \dx s^2=-c^2\dx t^2+a^2\dx\mathbf{r}^2.
\end{equation}
The energy-momentum tensor $T_{00}=\rho c^4$, $T_{0i}=0$, and $T_{ij}=\delta_{ij}pa^2$. We assume
\begin{equation}\label{eq:12}
  U_{\mu\nu}={\rm diag}(f_1c^2,f_2a^2,f_2a^2,f_2a^2),
\end{equation}
where $f_i=f_i(t)$. Substituting metric (\ref{eq:11}) and Eq. (\ref{eq:12}) into $\Box U_{\mu\nu}=R_{\mu\nu}$, we obtain
\begin{align}
  -\ddot{f}_1-3H\dot{f}_1+6H^2f_1+6H^2f_2&=-3\frac{\ddot{a}}{a},\label{eq:13} \\
  -\ddot{f}_2-3H\dot{f}_2+2H^2f_2+2H^2f_1&=2H^2+\frac{\ddot{a}}{a},\label{eq:14}
\end{align}
where $\dot{}\equiv\dx/\dx t$ and $H\equiv\dot{a}/a$. The above equations show $f_i$ is dimensionless. Substituting metric (\ref{eq:11}) and Eq. (\ref{eq:12}) into Eq. (\ref{eq:03}), we obtain
\begin{widetext}
\begin{align}
  &3\alpha_1H^2+\alpha_2\left(2\ddot{f}_1-6\ddot{f}_2-12H^2+\frac{9}{2}\dot{f}_2^2+\frac{1}{2}\dot{f}_1^2
  -12\frac{\ddot{a}}{a}-3\dot{f}_1\dot{f}_2-6H^2f_1+18H^2f_2\right)
  +\alpha_3\left(\frac{3}{2}\dot{f}_2^2+\frac{1}{2}\dot{f}_1^2+\ddot{f}_1\right.\nonumber\\
  &-3\frac{\ddot{a}}{a}-\frac{3}{2}Hf_1\dot{f}_1-\frac{9}{2}Hf_2\dot{f}_2
  +12H^2f_1f_2-12H^2f_1+\frac{3}{2}\frac{\ddot{a}}{a}f_1-\frac{3}{2}\frac{\ddot{a}}{a}f_2
  -\frac{1}{2}f_1\ddot{f}_1-\frac{3}{2}f_2\ddot{f}_2+3H\dot{f}_2+6H^2f^2_1\nonumber\\
  &\left.+6H^2f_2^2-9H^2f_2\right)=\zeta\rho c^4, \label{eq:15}\\
  &-\alpha_1\left(H^2+2\frac{\ddot{a}}{a}\right)+\alpha_2\left(-3\dot{f}_1\dot{f}_2+6H\dot{f}_2
  -2H\dot{f}_1+4\frac{\ddot{a}}{a}f_1-12\frac{\ddot{a}}{a}f_2+2H^2f_1-6H^2f_2+12H^2+\frac{9}{2}\dot{f}_2^2
  +\frac{1}{2}\dot{f}_1^2\right.\nonumber\\
  &\left.+12\frac{\ddot{a}}{a}\right)+\alpha_3\left(\frac{5}{2}\frac{\ddot{a}}{a}f_1
  +\frac{3}{2}\frac{\ddot{a}}{a}f_2-2\frac{\ddot{a}}{a}f_1^2-2\frac{\ddot{a}}{a}f_2^2
  +\frac{1}{2}f_1\ddot{f}_1+\frac{3}{2}f_2\ddot{f}_2+3H\dot{f}_2+4H\dot{f}_1-4H^2f_1^2-4H^2f_2^2+H^2f_2\right.\nonumber\\
  &\left.+2H^2+\frac{3}{2}\dot{f}_2^2+\frac{1}{2}\dot{f}_1^2+\ddot{f}_1+\frac{\ddot{a}}{a}
  -4Hf_1\dot{f}_2-\frac{5}{2}Hf_1\dot{f}_1+\frac{1}{2}Hf_2\dot{f}_2-4Hf_2\dot{f}_1
  -4\frac{\ddot{a}}{a}f_1f_2-8H^2f_1f_2\right)=\zeta p c^2. \label{eq:16}
\end{align}
\end{widetext}
The equation of state reads $p=w\rho c^2$, and $w=0$ for matter, $w=1/3$ for radiation. These five equations form a complete set of ordinary differential equations. One can directly verify the energy conservation equation
\begin{equation}\label{eq:17}
  \dot{\rho}+3H(\rho+p/c^2)=0
\end{equation}
can be derived from Eqs. (\ref{eq:13})--(\ref{eq:16}). In addition, substituting Eqs. (\ref{eq:13})--(\ref{eq:14}) into Eq. (\ref{eq:15}) to eliminate $\ddot{f}_i$ and setting $\alpha_2=-\alpha_3/2$, we obtain
\begin{align}\label{eq:18}
  3\alpha_1H^2&+\alpha_3\left[3H^2(f_1+f_2)^2-3H^2(f_1+3f_2)\right.\nonumber\\
  &\left.-6H\dot{f}_2+\frac{1}{4}\dot{f}_1^2+\frac{3}{2}\dot{f}_1\dot{f}_2-\frac{3}{4}\dot{f}_2^2\right]=\zeta\rho c^4.
\end{align}
Values of $\{f_1,\dot{f}_1,f_2,\dot{f}_2\}$ at one time point can be regarded as boundary conditions. Equation (\ref{eq:18}) imposes limits on these boundary conditions because $\rho$ should be positive. We can use Eqs. (\ref{eq:17})--(\ref{eq:18}) instead of Eqs. (\ref{eq:15})--(\ref{eq:16}) to study the evolution of the Universe. Taking the derivative of Eq. (\ref{eq:18}) with respect to $t$, and substituting Eqs. (\ref{eq:13}), (\ref{eq:14}), and (\ref{eq:17}) into the result to eliminate $\ddot{f}_i$ and $\dot{\rho}$, we obtain
\begin{align}\label{eq:19}
  \frac{\ddot{a}}{a}&=\frac{1}{\alpha_3[(f_1+f_2)^2-f_1-3f_2+1]+\alpha_1}\cdot
  \left[-\frac{1+w}{2}\zeta\rho c^4\right.\nonumber\\
  &\quad+\alpha_3H^2\left(f_1^2+f_2^2+2f_1f_2+f_1-f_2-2\right)\nonumber\\
  &\quad+\alpha_3H\left(\dot{f}_1-3\dot{f}_2-2f_1\dot{f}_1-2f_1\dot{f}_2-2f_2\dot{f}_1-2f_2\dot{f}_2\right)\nonumber\\
  &\quad\left.+\alpha_3\left(\frac{3}{2}\dot{f}_1\dot{f}_2+\frac{1}{4}\dot{f}_1^2-\frac{3}{4}\dot{f}_2^2\right)
  +\alpha_1H^2\right].
\end{align}

In order to explain the late-time acceleration, we expect a matter-dominated universe owns accelerating expansion solutions. Equation (\ref{eq:19}) meets our expectations. For example, in the case $\{\alpha_1=0,\alpha_3=1\}$, setting $\{H=H_0,f_1=0,\dot{f}_1=2H_0,f_2=0,\dot{f}_2=0\}$ at $t=t_0$ gives $\ddot{a}/a=H_0^2/2>0$ and $\rho=H_0^2/(8\pi G)>0$.

For the radiation-dominated universe, we know the horizon problem disappears if the expansion is accelerating. Equation (\ref{eq:19}) shows the radiation-dominated universe also has accelerating expansion solutions. For example, in the case $\{\alpha_1=0,\alpha_3=1\}$, setting $\{H=H_0,f_1=0,\dot{f}_1=2H_0,f_2=0,\dot{f}_2=0\}$ at $t=t_0$ gives $\ddot{a}/a=H_0^2/3>0$ and $\rho=H_0^2/(8\pi G)>0$.

So far, we have proved both the matter and radiation-dominated universe can be accelerating with suitable boundary conditions, i.e., values of $\{f_1,\dot{f}_1,f_2,\dot{f}_2\}$. Observational data can constrain these values. In addition, we should explore physical origins of the desired boundary conditions. For simplicity and beauty, we suggest readers focus on the case $\{\alpha_1=0,\alpha_2=-1/2,\alpha_3=1,\zeta=8\pi G/c^4\}$.

\section{Conclusions}
In this paper, we propose a new nonlocal gravity theory, which combines the advantages of the pure fourth-order gravity \cite{Tian2018} and Deser-Woodard nonlocal gravity \cite{Deser2007}. This is a pure geometric theory with second-order differential field equations in the localized formalism, and no parameters related to $H_0$ appear in the Lagrangian. The theory recovers Newton's gravity and gives $\Psi=\Phi$ in the weak field limit. The speed of gravitational waves equals to $c$. For cosmology, accelerating expansion can be obtained with suitable boundary conditions in both the matter and radiation-dominated universe. Especially, the above good properties also exist in the simplest case $\{\alpha_1=0,\alpha_2=-1/2,\alpha_3=1,\zeta=8\pi G/c^4\}$.

Extension of our work is possible. Combined with the result in \cite{Tian2018}, we conjecture there should be a large class of operators $\widehat{O}$ that make the theory with the following Lagrangian:
\begin{equation}
  \mathcal{L}_G=\frac{1}{2\zeta}\left(R_{\mu\nu}\widehat{O}^{-1}R^{\mu\nu}-\frac{1}{2}R\widehat{O}^{-1}R\right)
\end{equation}
behaves well in local gravitational systems, gravitational waves, and cosmology.

\section*{Acknowledgements}
We are grateful to the referees for valuable comments. This work was supported by the National Natural Science Foundation of China under Grant No. 11633001.

%

\end{document}